\documentclass[epj]{svjour}
\usepackage{graphics}
\usepackage{graphicx}
\usepackage{bbm}
\usepackage{amsmath,amssymb}

\newcommand{\be}{\begin{equation}}
\newcommand{\ee}{\end{equation}}
\newcommand{\bea}{\begin{eqnarray}}
\newcommand{\eea}{\end{eqnarray}}

\begin{document}
\title{Many-electron transport in Aharonov-Bohm interferometers:\\
A time-dependent density-functional study}
\author{A. Salman\inst{1,2}, V. Kotim\"aki\inst{2}, A. Siddiki\inst{3} and E. R\"as\"anen\inst{4,2}
}                     
%
%
\institute{Physics Department, Akdeniz University, Antalya, Turkey \and
Nanoscience Center, Department of Physics, University of Jyv{\"a}skyl{\"a}, FI-40014 Jyv{\"a}skyl{\"a}, Finland \and
Physics Department, Faculty of Sciences, Istanbul University, 34134 Vezneciler-Istanbul, Turkey \and
Department of Physics, Tampere University of Technology, FI-33101 Tampere, Finland}
\date{Received: date / Revised version: date}
%
\abstract{
We apply time-dependent density-functional theory to study many-electron
transport in Aharonov-Bohm interferometers in a non-equilibrium situation.
The conductance properties
in the system are complex and depend on the
enclosed magnetic flux in the
interferometer, the number of interacting particles, and
the mutual distance of the transport channels at the points of encounter.
Generally, the electron-electron interactions do not
suppress the visibility of Aharonov-Bohm oscillations if the
interchannel distance -- determined by the positioning of the
incompressible strips through the external magnetic field -- is optimized.
However, the interactions also impose an interesting
Aharonov-Bohm phase shift with channel distances below
or above the optimal one. This effect is combined with
suppressed oscillation amplitudes.
We analyze these effects within different approximations
for the exchange-correlation potential in time-dependent
density-functional theory.
}

\authorrunning{A. Salman {\em et al.}}
\titlerunning{Many-electron transport in AB interferometers}

\maketitle
\section{Introduction}

Integer and fractional quantum Hall effects~\cite{qh} together
with the Aharonov-Bohm (AB) effect~\cite{ab} enable
the design of interferometry devices that can be
controlled by an external magnetic field alone.
Significant experimental efforts in
two-dimensional (2D) systems have been made~\cite{Heiblum05:abinter,Goldman05:155313,Nissim09:,godfrey:07,Bernd:ABosc.,Marcus09}
to exploit the AB effect in the control of electron dynamics
in quantum Hall interferometers. The complexity of these
systems arises from the fact that not only the quantum mechanical
phase of the electrons, but their path itself, is affected
by the magnetic flux. Hence, the electron transport in
AB interferometers also provides a theoretical challenge
that should be approached beyond the single-particle edge-state
scheme~\cite{Buettiker86:1761} that does not take the path-dependence
completely into account~\cite{bernd:07}.

The characteristics of the measured AB oscillations have
been analyzed with edge-channel
experiments~\cite{Goldman05:155313,Neder06:016804}
and simulations~\cite{igor08:ab}.
In a recent work of some of the present authors~\cite{njp}
the picture was made more complete by combining
full electrostatics~\cite{Afif:AB} of a AB interferometer
device~\cite{Goldman05:155313} with a time-propagation of noninteracting
excess electrons in the obtained incompressible strips in
non-equilibrium. Clear AB oscillations
were found as a function of the magnetic flux, and the
visibility (amplitude) of the oscillations was found to
largely depend on the width and position of the current
channels.

In this work we extend a previous single-electron study~\cite{njp}
to the many-electron regime in the framework of time-dependent
density-functional theory~\cite{ullrich,Book-Gross} (TDDFT).
We consider different
(albeit adiabatic) approximations for the exchange-correlation
functional including the exact exchange (EXX), Hartree, and
local-density approximation (LDA). We find that in the
high-visibility regime, corresponding to an optimal positioning
of the incompressible strips (and thus the current channels),
the electron-electron interactions do not suppress the AB
oscillations. This is in accordance with a previous TDDFT study
on semiconductor quantum rings~\cite{ville}. However,
the interactions introduce an AB phase shift whose magnitude
is an increasing function of the electron number varied between
one and ten. In addition, the interactions suppress the
AB amplitudes if the current channels are either too close
or too far apart from each other. The induced phase shift
could be measured in an experiment, which would eventually
quantify the actual role of electronic interactions in a real setup.

\section{Model and methods}

\begin{figure}\vspace{0.0cm}
\includegraphics[width=0.99\columnwidth]{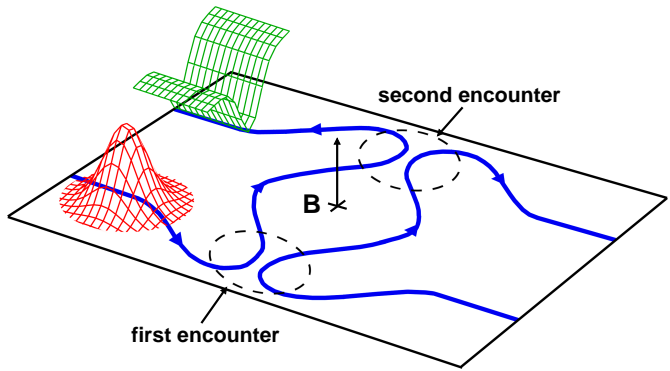}
\caption{(Color online) Schematic figure of the transport setup.
The upper mesh represents the cross
section of the model channel potential along the blue (black) thick
lines. The lower mesh shows the initial wave packet used 
in the time-propagation.}
\label{fig1}
\end{figure}

The transport simulation is carried out by injecting many-electron
wave packets into a static 2D potential.
Figure~\ref{fig1} shows a schematic picture of the system.
The shape of the model potential is based on the experimental
AlGaAs/GaAs device by
Camino \emph{et al.}\cite{Goldman05:155313} and electrostatic
calculations performed in Ref.~\cite{njp}. In particular,
the potential corresponds to incompressible strips with an
integer filling factor (here $\nu=2$) that carry the
non-equilibrium electric
current.
The model potential alongside the thick curves in Fig.~\ref{fig1} has
a Gaussian cross section, $V_{\rm cross}=-V_0 e^{-s^2/c^2}$, where $s$ is the
coordinate perpendicular to the channel, $V_0=80$ (Hartree
atomic units used throughout) is the channel
depth set to prevent the electrons from escaping the tracks, and
$c=0.3$ is the track width parameter. 
A part of the potential
is visualized as a mesh in the upper part of Fig.~\ref{fig1}.
A Gaussian cross section is a convenient choice for two reasons. First,
it is a good approximation for the magnetic (parabolic) confinement
at the bottom of the channel. Secondly, the upper part of
the potential allows ``leaking'' of the electron flow in consistency
with the experimental situation. It is important to note that
in the time-propagation we deal with the {\em excess electron density}
on top of the Fermi background that is assumed to be static.

We also apply a linear ramp potential to accelerate the initial
state, thus mimicking the source-drain voltage.
The initial state in the calculation is an $N$-electron wave packet at the
lower-left corner of the system (see the lower mesh in Fig.~\ref{fig1}).
It is calculated by solving the ground-state problem for 
a part of the input channel (confined by step functions) 
that contains $N$ Coulomb-interacting particles~\cite{ville}.

Our transport scheme described above leads to inevitable
{\em non-chiral} transport, i.e., the direction of the current
is the same on the both sides of the interferometer.
This is in accordance with {\em non-equilibrium} transport
experiments~\cite{newref}, where an external current is imposed and a
Hall (electrochemical) potential difference develops
at two opposite edges {\em with the same slope}~\cite{Guven03:115327}.
Such potential drops in the incompressible strip have been found
in several experiments~\cite{Ahlswede01:562,Ahlswede02:165,Dahlem:10}.
In the conventional equilibrium case, however, the current
would be chiral as confirmed in several studies.

In the time propagation of electrons we apply
TDDFT~\cite{ullrich,Book-Gross} on a 2D real-space grid. We use the
fourth-order Taylor expansion of the time evolution operator to the Kohn-Sham
states and perform the calculation with the OCTOPUS code package~\cite{octopus}.
The exchange-correlation potential in TDDFT is treated within adiabatic
LDA, and in the two-particle case also with
the (adiabatic) exact exchange approximation. In comparison, we have applied
the bare Hartree approximation (including only classical electrostatic
interactions and neglecting exchange and correlation) and
independent-electron approximation (neglecting interactions completely).

We consider the conductance of the system as a function of the uniform and
perpendicular magnetic-field strength in the units of the
enclosed flux quanta, $\Phi/\Phi_0$, where $\Phi=B\times {\rm area}$
and $\Phi_0=h/e$. Following the most experimental setups, the magnetic
field is homogeneously distributed across the whole system.
Another changing variable is the distance between
the points of encounter (interchannel distance) $d$
(see Fig.~\ref{fig1}) and the number of injected particles $N=$ 2, 6, or 10.
We consider only spin-compensated (zero total spin) systems.
We point out that in an experimental situation $d$ would be essentially
controlled either with the gates, or with the magnetic field that affects
the position and the width of the incompressible strips. 

We compute the area of the interferometer by integrating the
region enclosed by the curve drawn on the plane by the minima
of $V_{\rm cross}$, and cut at the points of encounter
(see Fig.~\ref{fig1}). Therefore, varying interchannel distance 
$d$ is taken into account 
in the calculation of the flux. However, the magnetic field 
affects only the flux but {\em not} the shape of the channels. 
This is justified by the 
following consideration. In a typical situation, notable changes 
in the spatial distribution of the incompressible strips can be 
seen in the scale of $\Delta B \sim 0.1\ldots 0.4$ T 
(see Fig. 2 in Ref.~\cite{njp}). However, with the dimensions 
of a typical sample in a micrometer range, an increase (decrease)
of the field strength by this amount would add (remove) 
about $100\ldots 400$ flux quanta. Therefore, as we 
consider changes within only three flux quanta, the shape of the 
channels can be kept fixed by a good approximation.

The relative conductance is
estimated as the number of electrons found in the top-right corner of the
system as a function of time~\cite{ville}.
Since the simulation box is finite, we monitor the electron
density only until back-scattering effects from the borders of the box appear.
The measured conductance parameter is the normalized probability density found
in the upper right corner of the system. We point out that the
setup is not symmetric with respect to the electron current; an
opposite direction of the magnetic field (${\bf B}\rightarrow -{\bf B}$)
would yield different results. Thus the situation 
is different from a typical symmetric transport experiment with a single
input and a single output lead. A more detailed discussion of
the model system can be found in Ref.~\cite{njp}.

\section{Results}

\begin{figure}
\includegraphics[width=0.95\columnwidth]{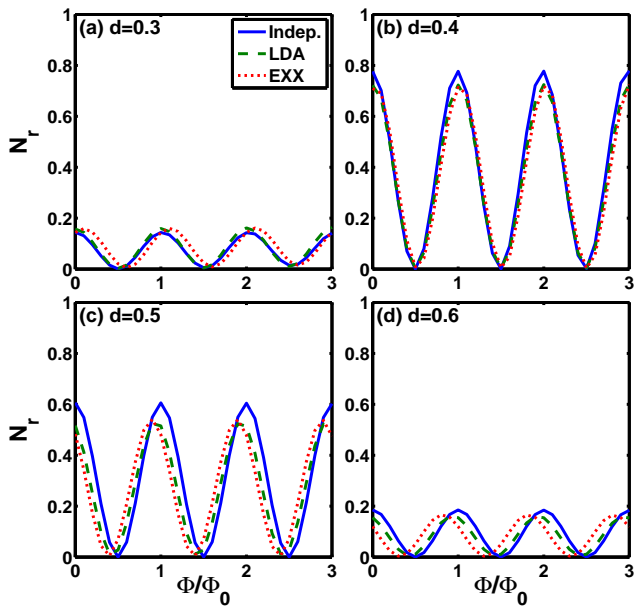}
\caption{(Color online) Estimated conductance in two-electron
Aharonov-Bohm interferometers with different
interchannel distances and different approximations
within time-dependent density-functional theory
(independent electrons, adiabatic local-density
approximation, adiabatic exact-exchange calculation).
The maximum visibility is obtained with $d\sim 0.4$.
}
\label{fig2}
\end{figure}

First we consider electron transport in a
two-electron system. The initial state is injected in the interferometer
using the linear ramp potential, and the relative conductance
is assessed as described in the previous section.
The panels in Fig.~\ref{fig2} show the conductance for different
interchannel distances, respectively, and different
approximations for the electron-electron interactions
are compared.

In all cases we find distinctive AB oscillations with
the correct flux quantization in the noninteracting limit. Hence,
the direction of the electron current (to the left or right drain)
is completely controlled by the flux. The maximum
amplitude (visibility) of the AB oscillations strongly depends
on the interchannel
distance. Regardless of the interactions, the maximum visibility
is found at $d\sim 0.4$. A detailed look in the electron densities
show that in this case the partitioning of the wave packet
in the first point of encounter (see Fig.~\ref{fig1}) is close to even,
i.e., one electron remains on the left track, but the other one
is transported to the right track. Consequently, the interference at
the second point of encounter is as complete as possible.

For two electrons the interactions seem to have a
minor effect on transport characteristics. They only
induce a small phase shift in the AB oscillations. The
direction of the phase shift depends on $d$: at $d\lesssim$ 0.4
(below the maximum visibility) the oscillations are shifted
to higher $\Phi/\Phi_0$ due to interactions, and at $d\gtrsim$ 0.4
the situation is opposite. The phase shifts are analyzed
in more detail below within Fig.~\ref{fig6}.

Moreover, as seen in Fig.~\ref{fig2}, the EXX and LDA results
are generally very close to each other. As the electronic correlations
are expected to be relatively small, the obtained agreement validates
the accuracy of LDA (as in EXX the exchange part is exact).
In this respect, our dynamical results are in accordance with previous
density-functional studies where the (surprisingly) high accuracy of LDA
results have been pointed out -- even in complicated situations such
as quantum Hall droplets with fractional filling
factors such as $\nu=5/2$ (Refs.~\cite{ari,spindroplet,rogge})
and $\nu=1/3$ (Ref.~\cite{henri}).

\begin{figure}
\includegraphics[width=0.9\columnwidth]{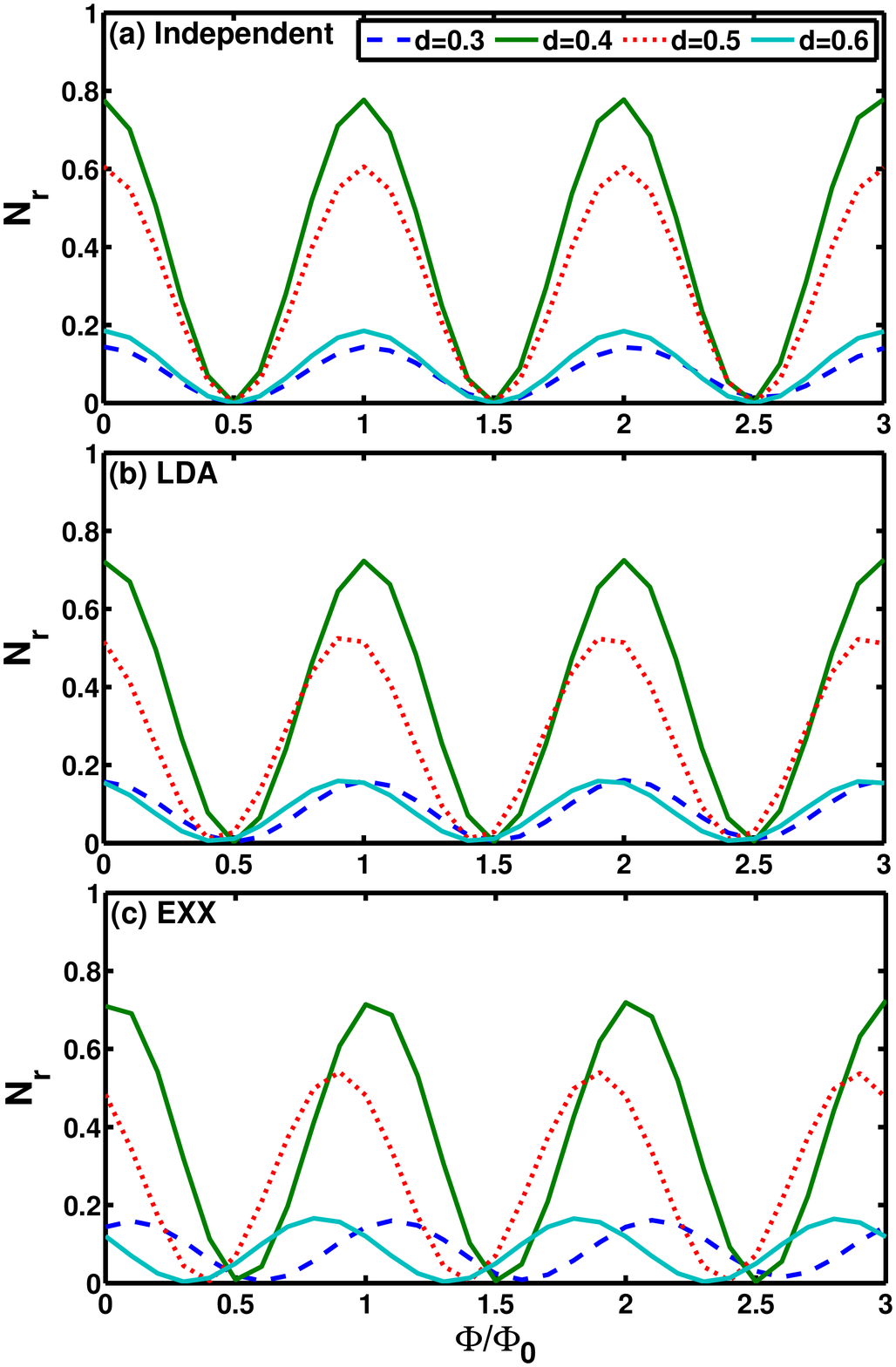}
\caption{(Color online) Same two-electron data as in Fig.~\ref{fig2} but
plotted separately for different approximations
within time-dependent density-functional theory.}
\label{fig3}
\end{figure}

In Fig.~\ref{fig3} we plot the same data as in Fig.~\ref{fig2}
but separate the results in terms of the interaction scheme.
The homogeneity of the AB oscillations in the noninteracting
case (a) is very clear, so that only the amplitude changes
as a function of $d$. As discussed above, LDA and EXX results
are similar and both show considerable phase shifts as
a function of $d$.

Next we consider the physical mechanism behind the phase
shift induced by interactions in Fig.~\ref{fig3}.
The Coulomb repulsion enhances the {\em spreading} of the wave
packet. This occurs in an asymmetric fashion, so that
there is relatively more spreading on the side of the
interferometer where the electron density is higher. As the spreading
is also radial, the outer parts of the wave packet
capture a larger phase than the inner parts. This effect
eventually leads to a shift in the AB oscillation phase when
the interactions are present.

\begin{figure}
\includegraphics[width=0.95\columnwidth]{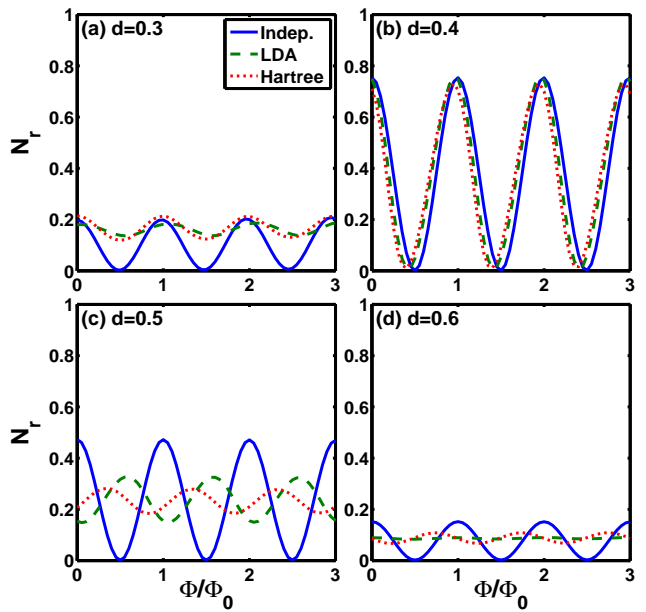}
\caption{(Color online) Estimated conductance in ten-electron
Aharonov-Bohm interferometers with different
interchannel distances and different approximations
within time-dependent density-functional theory
(independent electrons, adiabatic local-density
approximation, Hartree approximation).
}
\label{fig4}
\end{figure}

\begin{figure}
\includegraphics[width=0.9\columnwidth]{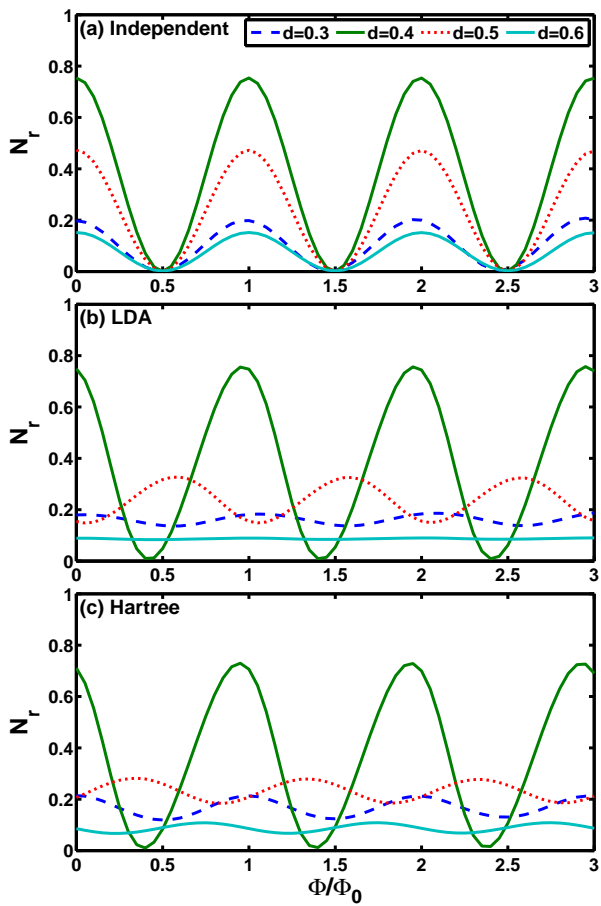}
\caption{(Color online) Same ten-electron data as in Fig.~\ref{fig4} but
now plotted separately for different approximations
within time-dependent density-functional theory.}
\label{fig5}
\end{figure}

\begin{figure}
\includegraphics[width=0.9\columnwidth]{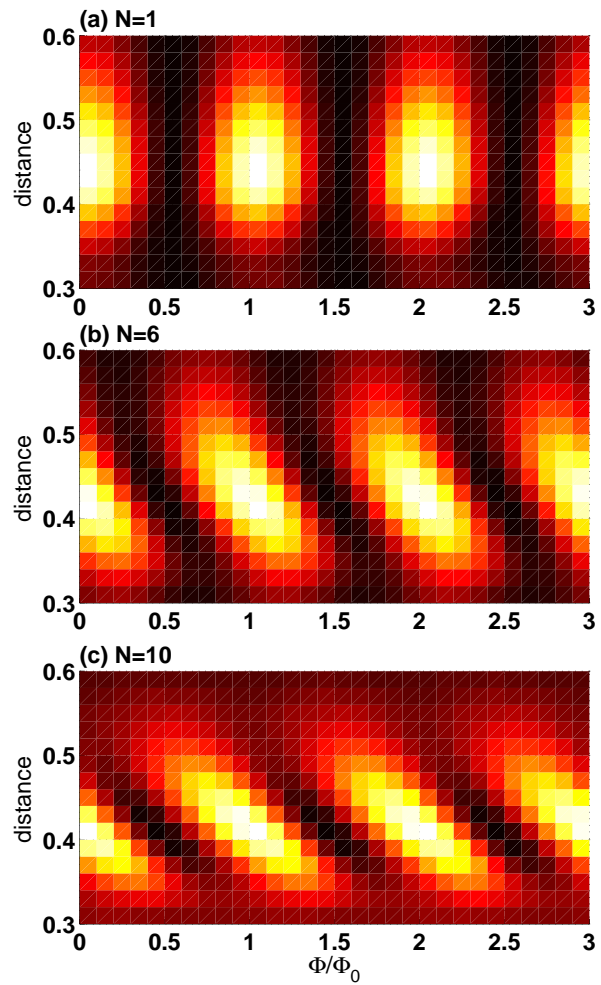}
\caption{(Color online) Aharonov-Bohm oscillations (color scale) for
one-, six-, and ten-electron systems, respectively.
Clear phase shifts can be in many-electron systems
treated here within adiabatic local-density
approximation.}
\label{fig6}
\end{figure}

In Fig.~\ref{fig4} we plot the conductance as a function
of the flux for a ten-electron system. The results with distinctive
AB oscillations are qualitatively similar to the two-electron situation
considered above. However, there are some important differences.
First, the maximum amplitude is obtained with a similar
interchannel distance ($d\sim 0.4$) as in the two-electron case.
This is due to the maximum partitioning ($50\%$ vs. $50\%$)
to the left and right channels with this $d$, so that the
underlying mechanism is similar to the two-electron system.
Secondly, we find that apart from the $d=0.4$ case the amplitude
is significantly dampened by the electron-electron interactions.
Thirdly, as shown in Fig.~\ref{fig5}, also the AB phase is affected
by the interactions more than in the two-electron system.
This can be understood by the fact that the density cloud is
larger, so that its parts capture different phases.
Consequently, the overall amplitude is partly smeared out. It is noteworthy
that in the case of $d=0.6$ the AB oscillations are almost
completely dampened in the LDA solution [see Fig.~\ref{fig5}(b)].

EXX calculations are not numerically stable for ten-electron
systems considered in Figs.~\ref{fig4} and \ref{fig5}.
However, we compare the adiabatic LDA calculations with the Hartree
results, where the exchange-correlation effects are
omitted. The results in the high-amplitude regime are
similar, and also generally the deviations are relatively
modest. Overall, the high-amplitude (high-visibility) regime
is not considerably affected by the electron-electron interactions.
Therefore, we might expect that in a real experimental setup
the AB oscillations are always present if the partitioning
is optimal, which is eventually determined by the location
of the incompressible strips controlled by the initial
magnetic field. Moreover, the AB phase can be controlled
by changing the positioning of the incompressible strips;
however, the phase change is achieved only with the price
of suppressing the visibility.

\begin{figure}
\includegraphics[width=0.8\columnwidth]{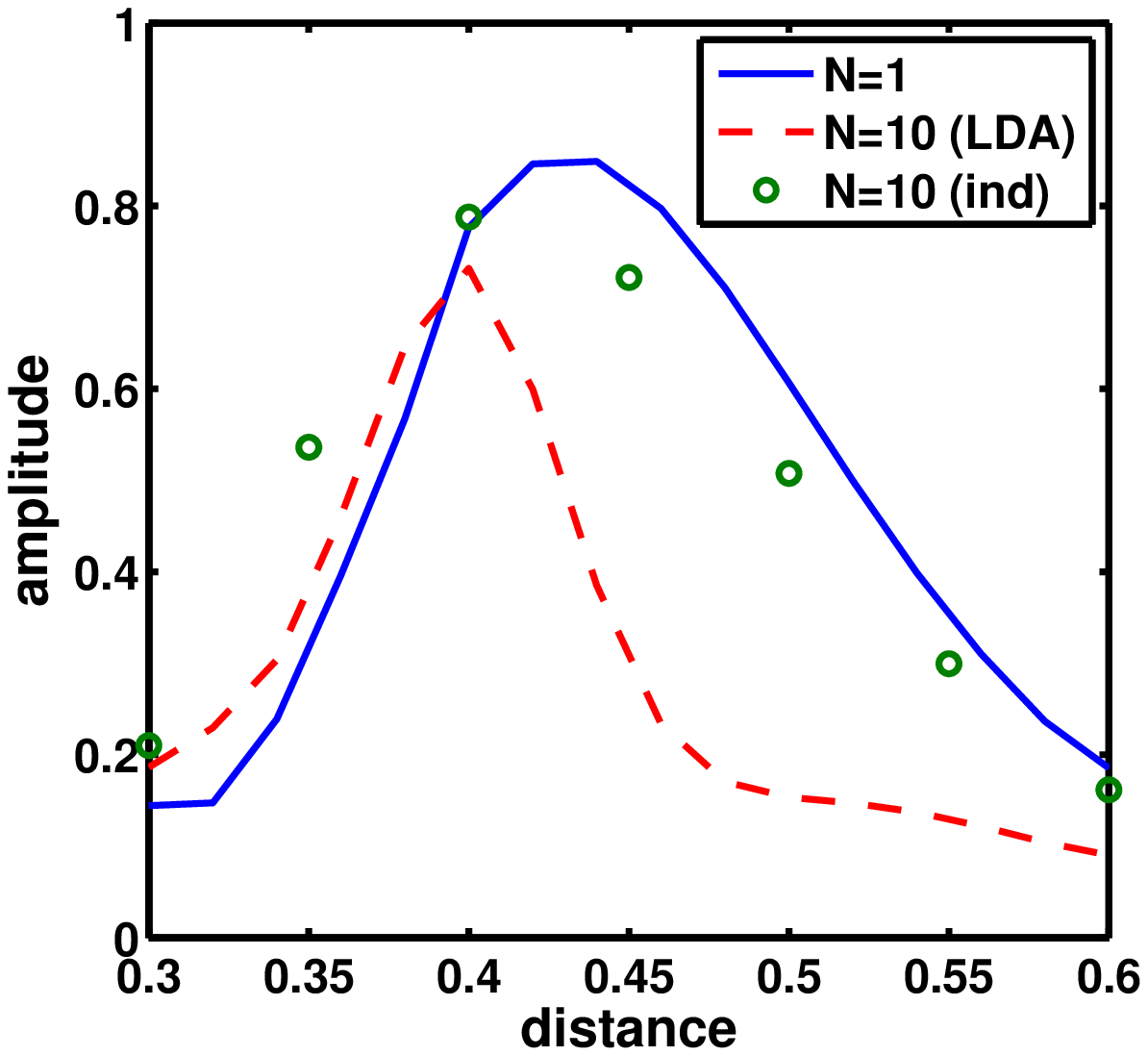}
\caption{(Color online) Maximum amplitude of the Aharonov-Bohm
oscillations (from minimum to maximum conductance)
in the range $\Phi/\Phi_0=0\ldots 3$ in the case of $N=1$ (solid line),
$N=10$ caluculated with the adiabatic local-density
approximation (dashed line) and without
interactions (circles).
}
\label{fig7}
\end{figure}

In Fig.~\ref{fig6} we present a summarized visualization
of the AB oscillation visibility with different
number of electrons propagating in the interferometer,
(a) $N=1$, (b) $N=6$, and (c) $N=10$. For the six- and ten-electron
results we use the adiabatic LDA. Our results demonstrate
that, as discussed above, the overall visibility is not reduced as a function of
$N$. 
Moreover, unless the interchannel distance is
close to the optimal value $d\sim 0.4$,
the interactions reduce the AB oscillation amplitudes.
This is visualized in Fig.~\ref{fig7} where we plot the
maximum amplitude in the whole range
$\Phi/\Phi_0=0\ldots 3$. We find that in the interacting $N=10$
case the amplitude decays close to zero at $d=0.6$, whereas
at that distance the single-electron result -- and
the $N=10$ without electron-electron interactions --
still have a $20\%$ visibility.

\section{Summary and discussion}

In summary, we have applied a single-electron propagation scheme
as well as time-dependent density functional theory to study
electron transport in realistic Aharanov-Bohm interferometers in
the integer filling factor regime. The device geometry has been
obtained from experimental boundary conditions that have been
applied in a self-consistent Poisson solver within the Thomas-Fermi
approximation in order to obtain the positions of the incompressible
strips. Here we have assumed, in accordance with previous experiments
and calculations, that the electric current is carried along the
incompressible strips in a non-chiral fashion (as the system is
in non-equilibrium). The current channels have been modeled with
a Gaussian tubes where electronic wave packets have been injected.

The time-evolution of excess electron density leads to distinctive
Aharonov-Bohm oscillations as a function of magnetic flux. The nature
of the oscillations depend on the electron-electron interactions
handled within (adiabatic) exact-exchange, local-density
approximation, and Hartree approximation, respectively,
in time-dependent density-functional theory. We have found
that the local-density approximation is accurate in comparison
with the exact-exchange approach.

The electron-electron interactions do not affect the maximum visibility
even if the number of propagating electrons is increased to ten. This
is in agreement with a previous time-dependent density-functional
study on many-electron semiconductor quantum rings~\cite{ville}.
However, we have found that around the optimal interchannel distance
the interactions induce a phase shift in the Aharonov-Bohm oscillations,
and the strength of the shift is proportional to the number of
electrons. If the interchannel distance is far from optimal the
interactions may dampen the Aharonov-Bohm oscillations completely.

Our results confirm that in a corresponding experimental setup
the Aharonov-Bohm oscillations should be clearly visible, especially
if the incompressible strips are sufficiently narrow. Furthermore,
possible phase shifts in the oscillations in different magnetic-field
ranges, i.e., with different positions and widths of the incompressible
strips, could reveal the importance of interaction effects in a
real setup. As a signature of such a dependency on
the interchannel distance we would like to mention the 
early experiments performed in Mach-Zehnder geometry~\cite{Neder06:016804}. 
There the visibility of Aharonov-Bohm oscillations is suppressed 
depending on the current excitation, and this phenomena is attributed 
to the energy dependence of the transmission probabilities. 
We hope that the present study encourages 
further experimental efforts in this direction.

\vspace{1cm}

This work was supported by T\"UB\.ITAK under grant no. 109T083,
Istanbul University (BAP:6970), Magnus Ehrnrooth Foundation,
ERASMUS Exchange Programme, Academy of Finland, and 
Institute of Theoretical and Applied Physics (ITAP) in
Turun\c c, Turkey. CSC Scientific Computing Ltd. is
acknowledged for computational resources.

%
%

%

\end{document}